\documentclass[twocolumn,preprintnumbers,superscriptaddress,amsmath,amssymb,showkeys]{revtex4}

\usepackage{graphicx}
\usepackage{dcolumn}
\usepackage{bm}
\usepackage{hyperref}
\usepackage{xcolor}
\usepackage{subfigure}
\usepackage{caption}

\begin{document}

\title{Modeling the Nervous System as An Open Quantum System}

\author{Yu-Juan Sun}
\affiliation{Department of Physics and Center for Quantum Information Science, National Cheng Kung University, Tainan 70101, Taiwan}
\author{Wei-Min Zhang}
\email{wzhang@mail.ncku.edu.tw}
\affiliation{Department of Physics and Center for Quantum Information Science, National Cheng Kung University, Tainan 70101, Taiwan}
\affiliation{Physics Division, National Center for Theoretical Sciences, Taipei 10617, Taiwan}

\date{\today}

\begin{abstract}
We propose a neural network model of multi-neuron interacting system that simulates neurons to interact each other through the surroundings of neuronal cell bodies. We physically model the neuronal cell surroundings, include the dendrites, the axons and the synapses as well as the surrounding glial cells, as a collection of all kinds of oscillating modes arisen from the electric circuital environment of neuronal action potentials. By analyzing the dynamics of this neural model through the master equation approach of open quantum systems, we investigate the collective behavior of neurons. After applying stimulations to the neural network, the neuronal collective state is activated and shows the action potential behavior. We find that this model can generate random neuron-neuron interactions and is proper to describe the process of information transmission in the nervous system physically, which may pave a potential route toward understanding the dynamics of nervous system.
\end{abstract}

\pacs{Valid PACS appear here}
\maketitle
\section{\label{sec:1}Introduction}
	The nervous system is a very complex part of living objects. It detects environmental changes that affect the body, and then cooperates with the endocrine system to respond to such events. It coordinates its movements and sensory information through signal transmission with different parts of its body \cite{tortora2018principles}. Therefore, understanding the structure and dynamics of the nervous system is the most challenging research field not only in biology, but also in the development of modern science and technology, especially for the development of deep learning methods in the machine learning framework, such as the computer vision task, speech recognition, multi-task prediction, and the autonomous vehicle, etc.~\cite{lecun2015deep, voulodimos2018deep, nassif2019speech, evgeniou2004regularized, xiao2020multimodal}. It may also help us to solve the complex nervous system and to construct proper implementations of the AI models associated with the understanding of the human brain \cite{hassabis2017neuroscience}. Meantime, investigating the nervous system can not only enable us to understand eventually the mechanism of human brain function, but also reveal the origin of human thought and consciousness \cite{turing1937computable, turing2009computing}.

	The key point of neural modeling is the connection between the biological evidence with the mathematical description through physical principles. Physically, we can depict all the phenomena into evolution of physical states for simplification. Each neuron is commonly described by two states, denoted as active and inactive states. The mathematical analogy of this concept was first proposed by McCulloch and Pitts in 1943 using logic circuit for describing the neural networks \cite{mcculloch1943logical}. About a decade later, Hodgkin and Huxley proposed the action potential model to explain the neuronal state activation through the experimental evidence of signal transmission from the neuronal membrane \cite{hodgkin1952quantitative}. They use electric circuits to mimic the lipid bilayer of membrane and the different kinds of ion channels through the membrane. This model can successfully fit the experimental data of the squid giant axon and other neural systems \cite{cole1939electric, hodgkin1949effect, keynes1951ionic, hodgkin1952quantitative, hodgkin1971conduction, bean2007action}. With the development of signal transmission and the binary neural state, the concept of neural network for brain modeling was also developed at almost the same time. Proposed in 1949, Hebb developed the theory to describe the formation of synapse \cite{hebb1949organization}. He found that when two joining neural cells fire simultaneously, the connection between them strengthens, which generates the new synapses. The concept of learning and memory explaining how the brain learns after firing several times through the connection between cells was developed \cite{hebb1949organization}. In 1982, Hopfield implemented Hebb's idea to simulate a neural network with a stochastic model, which is now called the Hopfield model. This model shows how the brain has the ability to learn and memorize \cite{hopfield1982neural}. In the Hopfield model, neurons are modeling as binary units or spins. The collective neurons form the state flow following the minimum ``energy" principle to reach the fixed points which are saved in advance. This has become the basic framework of the artificial neural network that AI development has relied in the past four decades.

	In the Hopfield model, neuron cells are randomly coupled to each other. It is known nowadays that the nervous signals transmit between neurons mainly via neurotransmitters. It is not clear how the nervous signals transmit through the randomly coupled neuron cells. In fact, the interactions between the neurons can be modeled with the surrounding including the neurotransmitters through the synapses, the electrical signal from the dendrites and the axons, and the noise from glial cells around the neurons \cite{sofroniew2010astrocytes}. Signal transmission via the intracellular wave propagation on the membrane has also been proposed from some evidence \cite{kaufmann1989ion, pollack2001cells, ivanova2003analyzing, heimburg2005soliton}. It is natural to ask how the neural surroundings affect the signal transmission between neurons, and it may be more fundamental to microscopically generate the random neuron couplings through these surrounding effects in neural modeling. As a result, the neural surroundings can be taken as an environmental effect with respect to the neural system in the nonequilibrium dynamics approach of open quantum systems, from which the dynamics of macroscopic collective phenomena of the neural network can be depicted.

	In the literature, there are some approaches describing the process of neuron dynamics from the nonequilibrium dynamics of open quantum systems, where the collective phenomena of neural networks are simulated as the spin collective phenomenon in open quantum systems determined by the master equation \cite{rose2016metastability, rotondo2018open, damanet2019atom, glauber1963time, peretto1984collective}. In the action potential picture, RC circuits are used to analog the neuronal axon membrane and various ion channels. However, some evidence of inductive features in the nervous system was observed \cite{cole1941longitudinal, wang2018inductance}. Thus the neural circuit for describing the signal transmission may also be an RLC circuit. RLC circuits are equivalent to a collection of coupled harmonic oscillators with various different frequencies \cite{feynman2000theory}, which gives us the idea to microscopically describe the environment effect of the nervous system as a continuous distribution of harmonic oscillators. On the other hand, the action potential is mainly due to the inward and outward ion flows of sodium and potassium ions through their own channels crossing the membrane, resulting in the voltage difference between the inside and outside of the membrane. The emergence of action potential transmission in the neurons is owing to the stimulation from the environment through the neurons coupling with the surroundings. We exert our neural network stimulations through the surrounding environment and show the collective neural state with the action potential behavior. In this work, we will focus on  investigating the collective behavior of the nervous system.

	The paper is organized as follows. In Sec.~\hyperref[sec:II]{\ref{sec:II}}, we introduce our neural model describing the interaction of the neurons in the neural network through the surroundings. The environment effects are modeled as a continuous distribution of harmonic oscillators with all kind of different frequencies. Then in Sec.~\ref{sec:III}, we derive the master equation which governs the dynamics of neurons, where the random couplings between neurons naturally emerge. In Sec.~\ref{sec:IV}, we analyze in detail the collective behavior of neurons in the neural networks. The collective neuron equations of motion are solved from the master equation obtained in Sec.~\ref{sec:III}. We stimulate the neural network from the environment, then the collective neural state shows the action potential properties. The thermal effects are also taken into account for mimicking the environment of the nervous system. A conclusion is made in Sec.~\ref{sec:V}.

\section{\label{sec:II} Modeling the nervous system as an open quantum system}

Inspired by previous neural modeling, such as the perceptrons \cite{rosenblatt1961principles}, the linear associative net \cite{cooper1979theory, longuet1968non, longuet1968holographic, kohonen2012associative}, and the content addressable memory \cite{amari1977neural, amari1978mathematical, little1974existence}, Hopfield modeled the neural network based on neuron statistical behavior \cite{hopfield1982neural}. He started with the binary neural states. The idea of binary neural state came from McCullough and Pitts's model \cite{mcculloch1943logical}, which characterizes the neural feature by "all-or-none" logical property. In Hopfield's model, the binary neural states can be realized through the spin configuration of $N$ spins at time $t$, 
\begin{align}
|\alpha, t\rangle = |\sigma^\alpha_1, \sigma^\alpha_2, ...,\sigma^\alpha_N; t\rangle,
\end{align} where $\alpha$ denotes different spin configurations and there are $2^N$ various configurations. The single spin state $\sigma^\alpha_i$ can be either 1 or -1 to represent the neuron being activating or inactivating respectively. The dynamical evolution of the neural states is determined through the coupling matrix $J_{N \times N}$. The state evolution from time $t$ to time $t'$ can be described with a matrix form of the transformation
\begin{align}
	\begin{bmatrix}
		\sigma_1^{t'} \\
		\sigma_2^{t'} \\
		\vdots \\
		\sigma_N^{t'} 
	\end{bmatrix}
	=
	\textbf{Sgn}\left\{\begin{bmatrix}
		J_{11} & J_{12} & \dots & J_{1N}\\
		J_{21} & J_{22} & \dots & J_{2N}\\
		\vdots & \vdots & \ddots \\
		J_{N1} & J_{N2} &  &J_{NN}
	\end{bmatrix}
	\begin{bmatrix}
		\sigma_1^t \\
		\sigma_2^t \\
		\vdots \\
		\sigma_N^t 
	\end{bmatrix}
	-
	\begin{bmatrix}
		b_1 \\
		b_2 \\
		\vdots \\
		b_N 
	\end{bmatrix}\right\}
\end{align} 
which follows the equation $\sigma_i^{t'} = \textbf{Sgn}\{\sum_j J_{ij}\sigma_j^t - b_i\}$, where the sign function $\textbf{Sgn}\{\}$ assigns the results 1 or -1 to the spin configuration elements with the time variable $t'$, and the threshold of the action potential for $i_{th}$ neuron represents by the bias $b_i$. The dynamical evolution leads the neural state to a local minimum in the configuration space. Thus, the Hopfield model can be equivalent to a disorder Ising model in statistics, in which neurons are modeled as spins and neuron dynamics is equivalently determined by an effective disorder Ising Hamiltonian:
\begin{eqnarray}
H_{\rm Hopfield} = \sum_{ij, i \neq  j}J_{ij}\sigma_i^z \sigma_j^z + h\sum_i\sigma_i^z.
\label{eq:2}
\end{eqnarray}
The first term in the above Hamiltonian describes the neuron coupling in the neural network, where $\sigma_i^z$ is the z-component of the spin Pauli matrix to represent two states of the silence (inactivate) and firing (activate) states of the $i_{th}$ neuron. The second term is an effective magnetic field, in response to the threshold of the action potential for firing. Such a Hamiltonian mimics the signal transmitting between the neurons, in which all neurons are connected to each other with randomly distributed couplings $J_{ij}$.  By defining the coupling through the concept from the Hebb’s learning theory, this neural network has the ability to learn and memorize what has saved. More specifically, the coupling strengths in Hopfield model define as $J_{ij} = \sum_{ij}\xi_i\xi_j$, which comes from the random variable {$\xi_i$} as a quenched, independent variable with the equal probability in 1 and -1 \cite{amit1985spin}. 

\begin{figure}[h!]
	\centering
	\subfigure{
		\begin{minipage}[t]{0.8\linewidth}
			\centering
			\includegraphics[width=\linewidth]{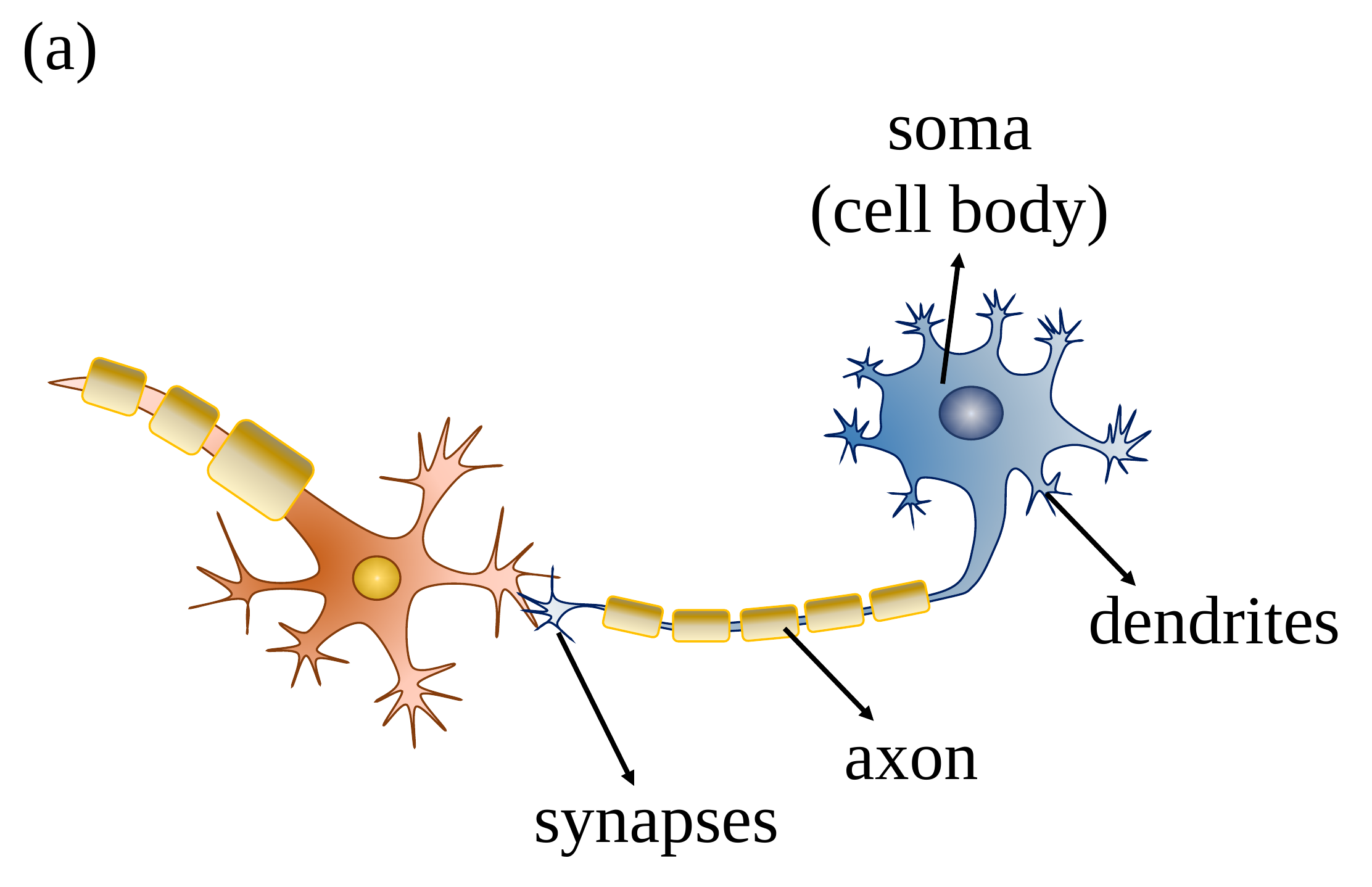}
			\label{fig:nervous}
		\end{minipage}%
	}%
	\quad
	\subfigure{
		\begin{minipage}[t]{0.8\linewidth}
			\centering
			\includegraphics[width=\linewidth]{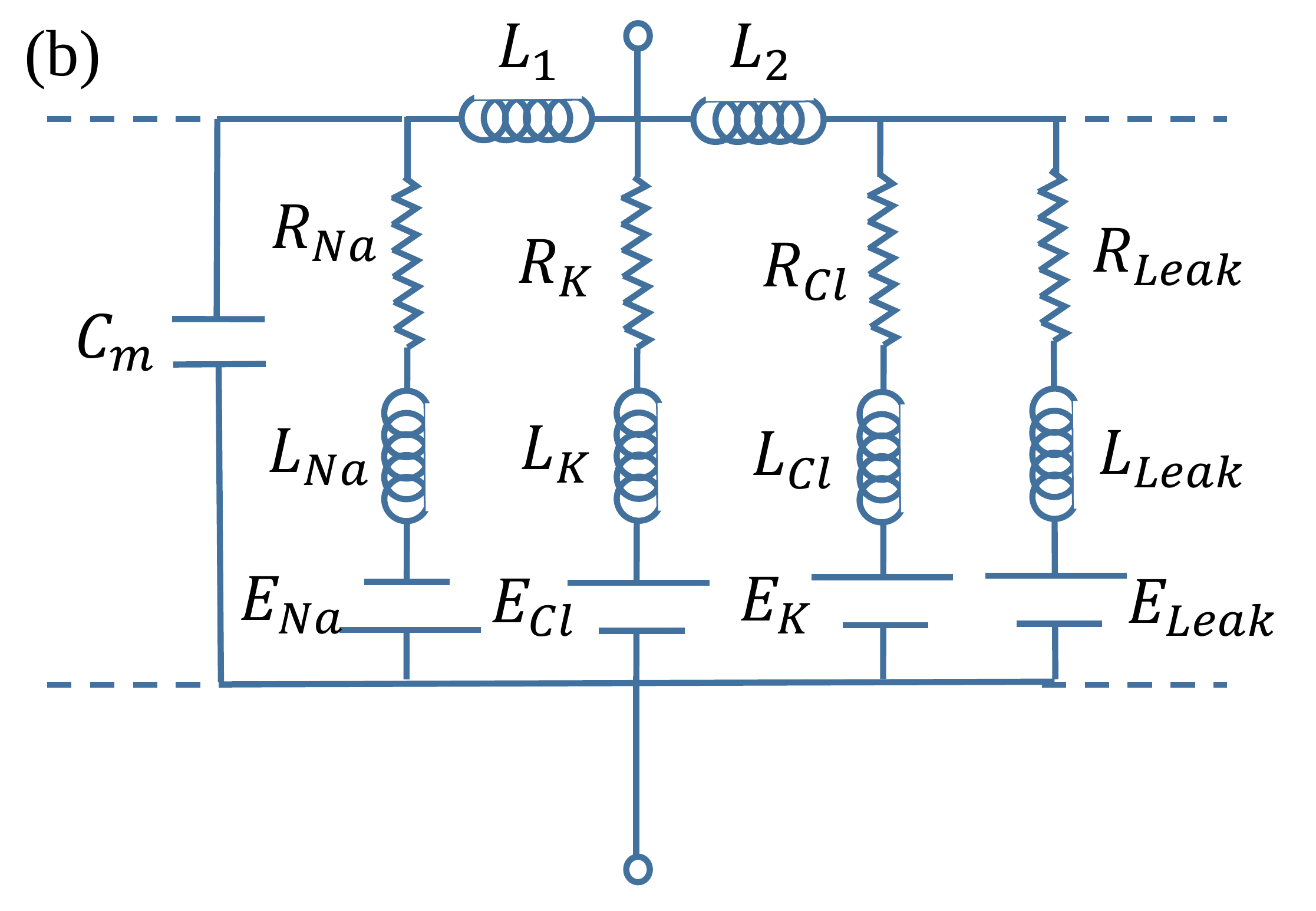}
			\label{fig:apmodel}
		\end{minipage}%
	}%
	\quad
	\subfigure{
		\begin{minipage}[t]{0.8\linewidth}
			\centering
			\includegraphics[width=\linewidth]{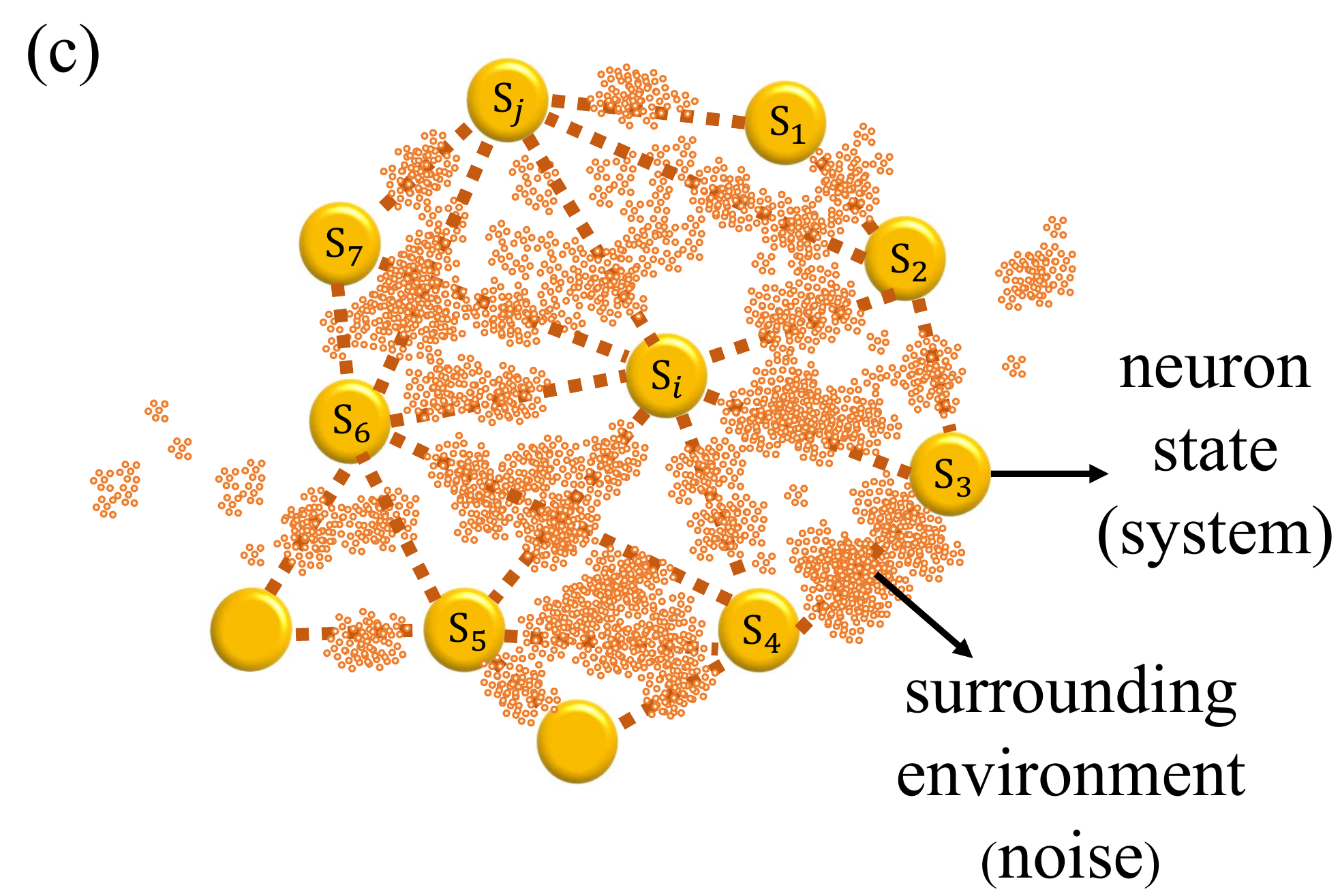}
			\label{fig:NNmodel}
		\end{minipage}
	}%
	\caption{\textbf{(a)} The schematic diagram to describe the portion of nervous system. \textbf{(b)} The schematic diagram for a modified action potential model to mimic the neural signal transmission with the complex damping circuit. \textbf{(c)} The schematic diagram to represent a set of neuronal cell bodies interacting to each other through the damping harmonic oscillators.}
	\label{fig:1}
\end{figure}

Our motivation is to find the microscopic picture of an equivalent random coupling for neuron-neuron interaction from the interactions between the neurons and their surroundings (environment). Physically, neuronal dynamics is governed by the interaction of neuronal cell body with their surroundings. The surrounding environment contains all matters surrounding the neuronal cell bodies, including axon, dendrite, synapses and the surrounding glial cells. On the other hand, the neural system transmits the electrical signal that one can measure. Hodgkin and Huxley use the cable model to explain the electric voltage change of the neuronal membrane through RC circuits. However, the circuit analogy of the nervous system should contain not only the resistance and capacitors from the foundation of the action potential model, but also the inductances as observed in some experiments \cite{cole1941longitudinal, wang2018inductance}. Consequently, the neural system can be taken as more reasonable RLC circuits modified from the action potential model, which shows in Fig.~\ref{fig:1}.
\captionsetup[figure]{format=plain,font=small,labelfont={color=black,bf},justification=raggedright}

In general, any circuit consisting of a complicated combination of RLC circuits corresponds to a collection of harmonic oscillators \cite{feynman2000theory}. Then, we can take the effect of all matters surrounding to the nervous system as a coupling of neurons with all possible harmonic oscillating frequencies, which is characterized in Fig.~\ref{fig:1}. In quantum mechanics, the collection of the harmonic oscillators with all possible different frequency can be expressed by the Hamiltonian $H_E =  \!\! \int \!\! \varrho(\omega) d\omega \hbar\omega a_\omega^\dag a_\omega$, where $a_\omega$ and $a_\omega^\dag$ are the annihilation and creation operators of the oscillating mode with the corresponding frequency $\omega$, $\varrho(\omega)$ is the density of state of all the oscillating modes. The coupling between neurons and all kinds of harmonic oscillator are given by the interaction Hamiltonian $H_{SE} =  \frac{\hbar}{\sqrt{N}} \int \varrho(\omega) d\omega \sum_{i} (V_{i}(\omega) a_\omega^\dag \sigma_i^- + V^*_{i}(\omega) \sigma_i^+ a_\omega) $,
where $V_{ik}$ is the coupling strength between the neuron $\sigma_i^z$ and the oscillating mode $\omega_k$. The spin up and spin down operators of each site are defined as $\sigma_i^\pm = \frac{1}{2}(\sigma_i^x + i\sigma_i^y)$, and $N$ is the total neuron number in the neural network. The neuron Hamiltonian $H_S=g\frac{\hbar}{2}\sum_i\sigma_i^z$ corresponds to a set of spins in an effective magnetic field $g$. 

	To sum up with the above model description, we can write down our neural system modeling by a neural network Hamiltonian, which contains three part, the Hamiltonians of the neurons, all oscillating modes from surroundings, and the couplings between them,
\begin{align}
H =&g \frac{\hbar}{2} \sum_i\sigma_i^z + \int \!\! \varrho(\omega) d\omega \hbar\omega a_\omega^\dag a_\omega \nonumber \\
&+  \int \!\! \varrho(\omega) d\omega \sum_{i} \frac{\hbar}{\sqrt{N}} (V_{i}(\omega) a_\omega^\dag \sigma_i^- + V^*_{i}(\omega) \sigma_i^+ a_\omega).
\label{eq:3}
\end{align}
We will show that it is the coupling between neurons and the harmonic oscillating modes of the neural surroundings bringing neurons connection to each other and leading to the random neuron-neuron coupling. 

\section{\label{sec:III} The master equation of the neuron open system dynamics}

The neural dynamics is described by the nonequilibrium evolution of the collection of all the neuronal states in the neural network. Neurons can interact to each other through their surrounding environment, as described by the Hamiltonian showing in Eq.~(\ref{eq:3}). Quantum mechanically, the total state evolution of neurons plus the environment is determined by the total density matrix $\rho_{\rm tot}(t)$ which carries all the state information of the neurons and their surroundings. It is governed by the Liouville-von Neumann equation \cite{von2018mathematical}, 
\begin{align}
\frac{d}{dt}\rho_{\rm tot}(t) = \frac{1}{i\hbar} [H, \rho_{\rm tot}(t)],  \label{vNe}
\end{align} 
where $H$ is the total Hamiltonian. We care about the state evolution of the neurons effected by the infinite number of oscillating modes in the environment considered in Eq.~(\ref{eq:3}). Thus, we should focus on the time evolution of the reduced density matrix $\rho_S(t)$ which is determined by tracing over all the environmental states from the total density matrix: 
$\rho_S(t)={\rm Tr}_E[\rho_{\rm tot}(t)]$. The reduced density matrix $\rho_S(t)$ describes the nonequilibrium dynamical evolution of neural states in the neural network. The equation of motion for the reduced density matrix $\rho_S(t)$ that determines such evolution is called the master equation which can be derived below. 

In the interacting picture of quantum mechanics, the total state of neurons plus their environment in the neural network is defined by $\widetilde{\rho}_{\rm tot}(t)= e^{iH_0t}\rho_{\rm tot}(t)e^{-iH_0t}$, where $H_0=H_S+H_E$. An expansion solution of Eq.~(\ref{vNe}) in the interacting picture can be written as
\begin{align}
\widetilde{\rho}_{\rm tot}(t)=&\widetilde{\rho}_{\rm tot}(0) + \frac{1}{i\hbar} \!\! \int_0^t \!\! dt'[H_I(t'), \widetilde{\rho}_{\rm tot}(t')] \notag \\
=& \widetilde{\rho}_{\rm tot}(0) + \frac{1}{i\hbar} \!\!\int_0^t\!\! dt'[H_I(t'), \widetilde{\rho}_{\rm tot}(0)] \notag \\
&-  \frac{1}{\hbar^2} \!\! \int_0^t\!\! dt' \!\!\int_0^{t'} \!\!dt''[H_I(t'), [H_I(t''),\widetilde{\rho}_{\rm tot}(t'')]],
\label{eq:5}
\end{align}
where the interacting Hamiltonian is given by $H_I(t)=e^{iH_0t}H_{SE}
e^{-iH_0t}$, and $\widetilde{\rho}_{\rm tot}(0)$ is the initial state. Supposing that the initial state of the system (neurons) and the environment (surroundings) is a decoupled state: $\widetilde{\rho}_{\rm tot}(0) = \widetilde{\rho}_S(0) \otimes \widetilde{\rho}_B(0)$, and the environment state is in a thermal equilibrium state, which is $\widetilde{\rho}_B(0)=\rho_B(0) = \frac{1}{Z_B} e^{-\beta H_E}$,
where $\beta = \frac{1}{k_B T}$ is the inverse temperature of the environment and $Z_B={\rm Tr}[e^{-\beta H_E}]$ is the environmental partition function. Meanwhile, we
assume that all the neurons and their surroundings are weakly coupled to each other so that the environmental state almost remains unchanged, namely, $\widetilde{\rho}_{\rm tot}(t'') \simeq \widetilde{\rho}_S(t'') \otimes \widetilde{\rho}_B(0)$ in Eq.~(\ref{eq:5}), which is called as the Born approximation. Because the neurons are weakly coupled to their environment, we can use further the Markov approximation, through replacing $\widetilde{\rho}_S(t'')$ in the above Born approximation by $\rho_S(t)$. After such Born and Markov approximations, we take the trace over all the environmental states. It can be shown that ${\rm Tr}_E[H_I(t), \widetilde{\rho}_{\rm tot}(0)] = 0$. Then we obtain the master equation for the reduced density matrix $\widetilde{\rho}_S(t)$ of all the neuron states,
\begin{equation}
\frac{d}{dt}\widetilde{\rho}_S(t) = -\frac{1}{\hbar^2} \!\! \int_{0}^{t}\!\! ds {\rm Tr}_E [H_I(t),[H_I(s), \widetilde{\rho}_S(t) \otimes \rho_B(0)]].
\label{eq:6}
\end{equation}
This is the Born-Markov master equation of open quantum systems \cite{breuer2002theory}.

Now, we apply the above master equation formulation to the Hamiltonian described by Eq.~(\ref{eq:3}). For simplicity, we assume that the coupling strength is a real value, $V_{i}(\omega) = V_{i}^*(\omega)$. In the Interacting picture, the interaction Hamiltonian is
\begin{align}
H_I(t) = \int \!\! \varrho(\omega) d\omega \sum_{i} \frac{\hbar V_{ik}}{\sqrt{N}} \big(& e^{i(\omega - g)t} \sigma_i^-a_\omega^\dag \notag \\
& + e^{-i(\omega - g)t} \sigma_i^+ a_\omega\big).
\label{eq:7}
\end{align}
After completing the trace over the environmental states in Eq. (\ref{eq:6}) and changing the formulation back into the Schr\"{o}dinger picture,we have the master equation with neuron degrees of freedom only
\begin{align}
\frac{d}{dt} &\rho_S(t)=\frac{1}{i\hbar}[{H_S'}(t), \rho_S(t)]  \notag \\
+ &\sum_{ij}  \Big\{\kappa_{ij}(t) \big[\sigma_j^- {\rho_S}(t) \sigma_i^+ \!-\! \frac{1}{2} \sigma_i^+ \sigma_j^- {\rho_S}(t) \!-\! \frac{1}{2} {\rho_S}(t) \sigma_i^+ \sigma_j^-\big]  \notag \\
&+\widetilde{\kappa}_{ij}(t) \big[\sigma_j^+ {\rho_S}(t) \sigma_i^- \!-\! \frac{1}{2} \sigma_i^- \sigma_j^+ {\rho_S}(t) \!-\! \frac{1}{2} {\rho_S}(t) \sigma_i^- \sigma_j^+ \big] \Big\}, \label{bmme}
\end{align}
which describes the dynamical evolution of all the neural states in the neuronal network. In Eq.~(\ref{bmme}), the first term in the right hand side is a unitary transformation for the system dynamics according to the renormalized system Hamiltonian ${H_S'}(t)=H_S + \delta H(t)$, where $\delta H(t)$ is induced by the couplings between the neurons and their surroundings 
\begin{eqnarray}
\delta H(t) = \sum_{ij} [ \lambda_{ij}(t)\sigma_i^+ \sigma_j^- - \widetilde{\lambda}_{ij}(t) \sigma_i^- \sigma_j^+ ],
\label{eq:8}
\end{eqnarray}
which characterizes how the disorder neuron-neuron interactions are generated from the coupling between the neurons and their surroundings. The other terms in the master equation describes the dissipation and fluctuations of the neurons. 

The environment-induced neuron-neuron interactions and the dissipation and fluctuation dynamics of neurons are determined by the corresponding time-correlations between the neurons and their surroundings in terms of the time-dependent coefficients in the master equation of Eq.~(\ref{bmme}),
\begin{subequations}
\label{eq:11}
\begin{align}
	&\kappa_{ij}(t) = \!\!\int_0^t \!\!ds\!\! \int_0^\infty \!\!\!\frac{d\omega}{2\pi} J_{ij}(\omega) (2\cos[(\omega\!-\!g)(t\!-\!s)])\bar{n}(\omega, T), \\
	&\widetilde{\kappa}_{ij}(t) = \!\!\int_0^t \!\!ds\!\! \int_0^\infty \!\!\!\frac{d\omega}{2\pi}J_{ij}(\omega)(2\cos[(\omega\!-\!g)(t\!-\!s)])n(\omega, T), \\
	&\lambda_{ij}(t) = -\!\!\int_0^t \!\!ds\!\! \int_0^\infty \!\!\! \frac{d\omega}{2\pi}J_{ij}(\omega) (\sin[(\omega\!-\!g)(t\!-\!s)])\bar{n}(\omega, T),\\
	&\widetilde{\lambda}_{ij}(t) = -\!\!\int_0^t \!\!ds\!\! \int_0^\infty \!\!\!\frac{d\omega}{2\pi} J_{ij}(\omega) (\sin[(\omega\!-\!g)(t\!-\!s)])n(\omega, T),
\end{align}
\end{subequations}
where $\kappa_{ij}$ and $\widetilde{\kappa}_{ij}$ describes the effects of environmentally induced dissipation and fluctuation,  $\lambda_{ij}$ and $\widetilde{\lambda}_{ij}$ are the environmentally induced random neuron-neuron couplings. The function $J_{ij}(\omega)$ is the spectral density of the environment
\begin{align}
J_{ij}(\omega) 
& = 2 \pi{\varrho}(\omega)V_i(\omega)V_j^*(\omega),
\label{eq:9}
\end{align}
and ${\varrho}(\omega)$ is the density of states of the environmental oscillating spectrum. The spectral density encompasses all the information about the structure of the materials surrounding neurons and the couplings with the neurons.
The function $n(\omega, T) = {\rm Tr}_E[\widetilde{\rho}_B(0)a_\omega^\dag a_\omega]$ is the particle distribution of the environmental oscillating modes, and $\bar{n}(\omega, T) =n(\omega, T)+1 $. 

The neural network contains plenty of dynamical neurons and it is difficult to solve it even numerically. We can lower the calculating cost by summing up all the neuronal operators as a collective neural spin. The collective neural spin $\hat{\vec{S}} = (\hat{S}^x, \hat{S}^y, \hat{S}^z)$ operator is defined by summing up all the neural spins in each direction $\alpha$
\begin{eqnarray}
\hat{S}^\alpha = \sum_i \sigma_i^\alpha,
\end{eqnarray}
where $\alpha = x, y, x$.

For formulating the collective neuron behavior conveniently, we assume that the environment provides the same effect on all the neurons, namely, the coupling strength being independent to different neurons, $V_i(\omega) = V(\omega)$. The spectral density then becomes to
\begin{eqnarray}
J_{ij}(\omega) \rightarrow J(\omega)  = 2\pi{\varrho}(\omega)|V(\omega)|^2.
\end{eqnarray}
As a result, the master equation is simply reduced to the form of
\begin{align}
\frac{d}{dt} & \rho_S(t) =\frac{1}{i\hbar}[{H_S'}(t), \rho_S(t)]  \notag \\
&+ \kappa(t) \Big[\hat{S}^- {\rho_S}(t) \hat{S}^+ 
\!-\! \frac{1}{2} \hat{S}^+ \hat{S}^- {\rho_S}(t) \!-\! \frac{1}{2} {\rho_S}(t) \hat{S}^+ \hat{S}^-\Big] \notag \\ 
&+ \widetilde{\kappa}(t)\big[ \hat{S}^+ {\rho_S}(t) \hat{S}^- \!-\! \frac{1}{2} \hat{S}^- \hat{S}^+ {\rho_S}(t) \!-\! \frac{1}{2} {\rho_S}(t) \hat{S}^- \hat{S}^+ \big],
\label{eq:me}
\end{align}
where $H_S'(t) = \lambda(t) \hat{S}^+ \hat{S}^- - \widetilde{\lambda}(t) \hat{S}^- \hat{S}^+$, and $\hat{S}^\pm = \frac{1}{2}(\hat{S}^x \pm i\hat{S}^y)$. The corresponding renormalization and dissipation/fluctuation coefficients in the master equation Eq.~(\ref{eq:me}) become
\begin{subequations}
\label{dfc}
	\begin{align}
		&\kappa(t) = \!\!\int_0^t\!\! ds \!\!\int_0^\infty \!\!\!\frac{d\omega}{2\pi} J(\omega) (2\cos[(\omega\!-\!g)(t\!-\!s)])\bar{n}(\omega, T),\label{eq:10} \\
		&\widetilde{\kappa}(t) = \!\!\int_0^t \!\!ds\!\! \int_0^\infty \!\!\!\frac{d\omega}{2\pi}J(\omega)(2\cos[(\omega\!-\!g)(t\!-\!s)])n(\omega, T),\\
		&\lambda(t) = -\!\!\int_0^t \!\!ds \!\!\int_0^\infty\!\!\! \frac{d\omega}{2\pi}J(\omega) (\sin[(\omega\!-\!g)(t\!-\!s)])\bar{n}(\omega, T),\\
		&\widetilde{\lambda}(t) = -\!\!\int_0^t \!\!ds\!\! \int_0^\infty\!\!\! \frac{d\omega}{2\pi} J(\omega) (\sin[(\omega\!-\!g)(t\!-\!s)])n(\omega, T).
	\end{align}
\end{subequations}
This is the master equation for the collective neural states of the nervous system.

\section{\label{sec:IV}Collective Behavior and Neural Dynamics Analysis}
\subsection{\label{sec:3_1}Equation of motion for the collective neural states}

The equation of motion for the collective neural states are obtained through the expectation values of the collective neural spin operators. By taking the expectation value of the collective spins with the reduced density matrix,
\begin{eqnarray}
S^\alpha= {\rm Tr}[\rho_S(t) \hat{S}^\alpha ] = \langle \hat{S}^\alpha \rangle,
\end{eqnarray}
and applying the mean-field approximation, $\langle \hat{S}^\alpha \hat{S}^\beta \rangle = \langle \hat{S}^\alpha \rangle \langle \hat{S}^\beta \rangle$, we obtain a close form of the equation of motion for the collective neural states from the master equation Eq.~(\ref{eq:me}).
\begin{subequations}
\label{eq:33}
	\begin{align}
	& \dot{S^x} = K(t){S^y}{S^z} \!-\! [g\! +\! P(t)]{S^y} \!+\! \frac{1}{2}[D(t){S^z} \!-\! F(t)]{S^x},
	\label{eq:31} \\
	& \dot{S^y} = - K(t){S^x}{S^z} \!+\! [g \!+\! P(t)]{S^x} \!+\! \frac{1}{2}[D(t){S^z} \!-\! F(t)]{S^y},
	\label{eq:32} \\
	&\dot{S^z} = -\frac{1}{2}D(t)({S^x}^2 \!+\! {S^y}^2) \!-\! F(t){S^z},
	\end{align}
\end{subequations}
where the coefficients in the above equation of motion are given by $K(t) = \lambda(t) - \widetilde{\lambda}(t), P(t) = \lambda(t) + \widetilde{\lambda}(t), D(t) = \kappa(t) - \widetilde{\kappa}(t), F(t) = \kappa(t) + \widetilde{\kappa}(t)$. The coefficients $K(t)$ and $P(t)$ are related with the neuron-neuron interactions, and $D(t)$ and $F(t)$ are related to the dissipation and fluctuations. All of them are induced by the couplings between neurons with the environment in the neural network. 

In the following calculation, as an example, we take the most common spectral density \cite{leggett1987dynamics}
\begin{equation}
J(\omega) = 2\pi\eta\omega (\frac{\omega}{\omega_c})^{s-1} e^{-\frac{\omega}{\omega_c}},
\label{eq:34}
\end{equation}
where $\eta$ is a dimensionless coupling constant between the system and the environment, and the $\omega_c$ is a cut-off frequency. The value of $s$ can be $s=1$, $<1$ and $>1$, corresponding to the spectrum of the environment being Ohmic, sub-Ohmic and super-Ohmic spectrum, respectively. Here we consider the Ohmic spectrum, $s=1$. With such detailed structure for the neuron environment, we can study the collective neural behavior induced by the environment effects.

\subsection{\label{sec:3_2}The dynamics of collective neural states}

\begin{figure}[h!]
	\centering
	\subfigure{
		\begin{minipage}[t]{0.8\linewidth}
			\centering
			\includegraphics[width=\linewidth]{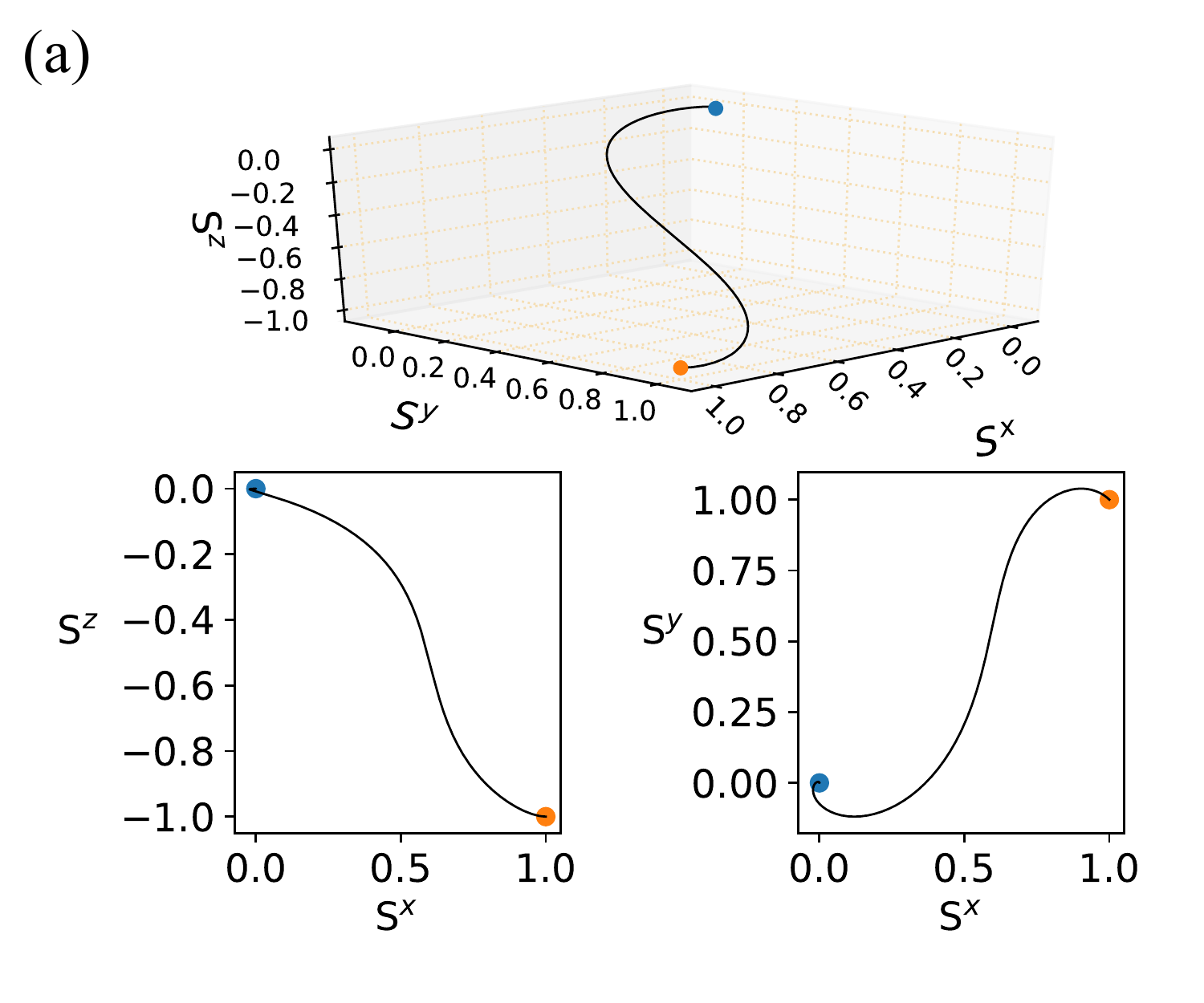}
			\label{fig:2a}
		\end{minipage}%
	}%
	\quad
	\subfigure{
		\begin{minipage}[t]{0.8\linewidth}
			\centering
			\includegraphics[width=\linewidth]{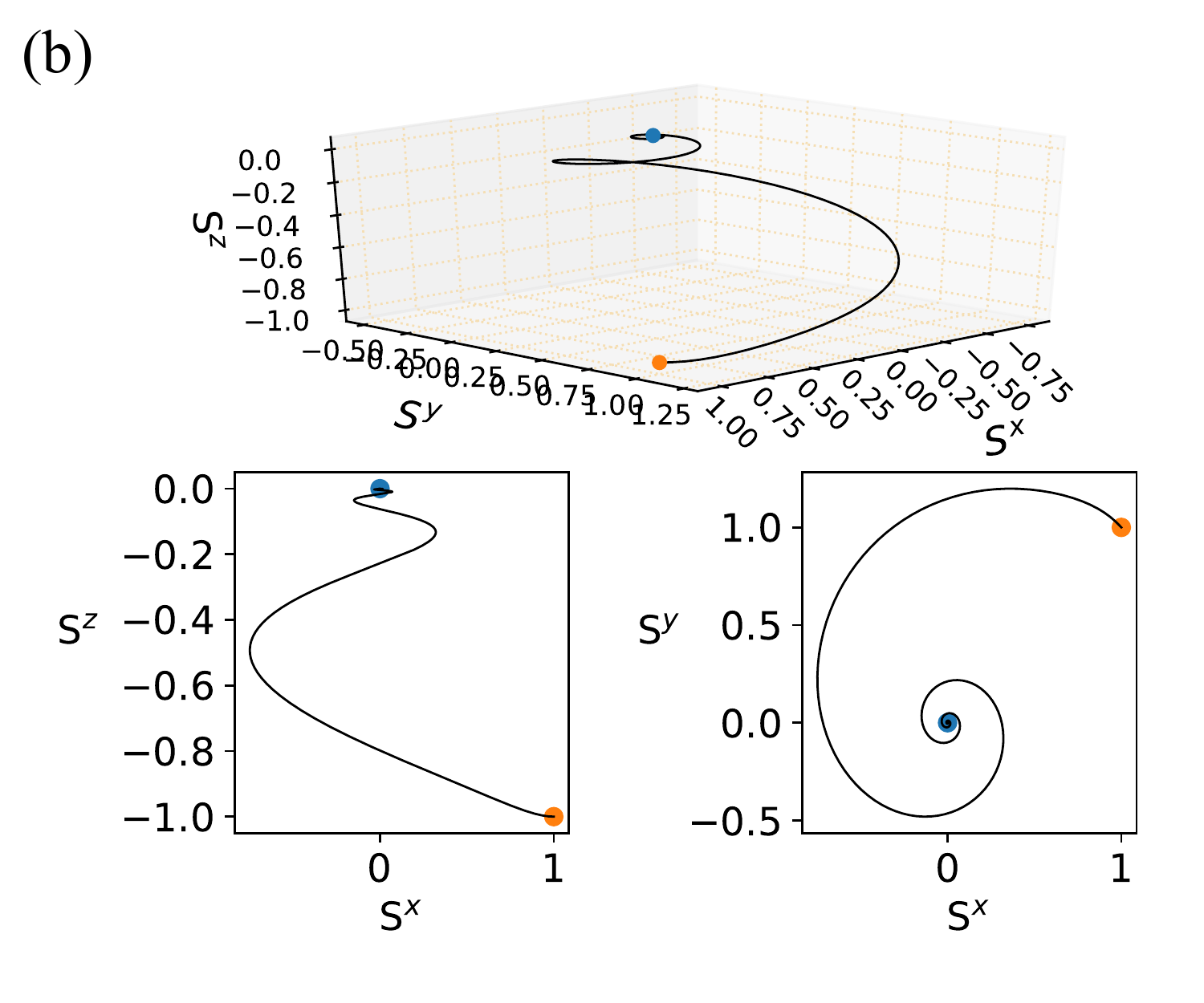}
			\label{fig:2b}
		\end{minipage}%
	}%
	\caption{Collective state evolution of the neural system coupled to the environment effect with different couplings. The couplings $\eta$ are \textbf{(a)}. $\eta = 0.1$ and \textbf{(b)}. $\eta = 0.02$. The temperature is taken in $T=300 K$ (room temperature), where $g = \omega_0$ determines the system frequency $\omega_0 = 2000$ Hz.}
	\label{fig:2}
\end{figure}

With the equation of motion Eq.~(\ref{eq:33}) obtained from the master equation, we want to explore how the collective neural states evolve in time under the coupling with their surroundings. We start from the collective neural states $(S^x, S^y, S^x) = (1, 1, -1)$ and study the time evolution of the collective state under the different coupling constant. The results are presented in Fig.~\ref{fig:2} with the coupling strength being $\eta= 0.1$ and $\eta= 0.02$. As one can see, the system always flows to the (0,0,0) state, which corresponds to the depolarized state. The differences are manifested in the trajectories of the collective neural states in the phase space. The larger of the coupling constant the sooner of the collective neural state is depolarized.
\captionsetup[figure]{format=plain,font=small,labelfont={color=black,bf},justification=raggedright}

In reality, the neural signal transmissions through the external pulses stimulate the neurons. In the following, we want to explore how the collective neural states evolve in time when exerting an external pulse to the network by replacing the constant coupling strength with a rectangular pulse. To investigate the collective neural dynamics, we consider first the simple case of the nervous system at zero temperature and then move to the more realistic case of the nervous system at room temperature. 

\subsubsection{\label{sec:3_2_1}Collective neural dynamics at zero temperature.}

\begin{figure}[htbp]
	\centering
	\includegraphics[width=1\linewidth]{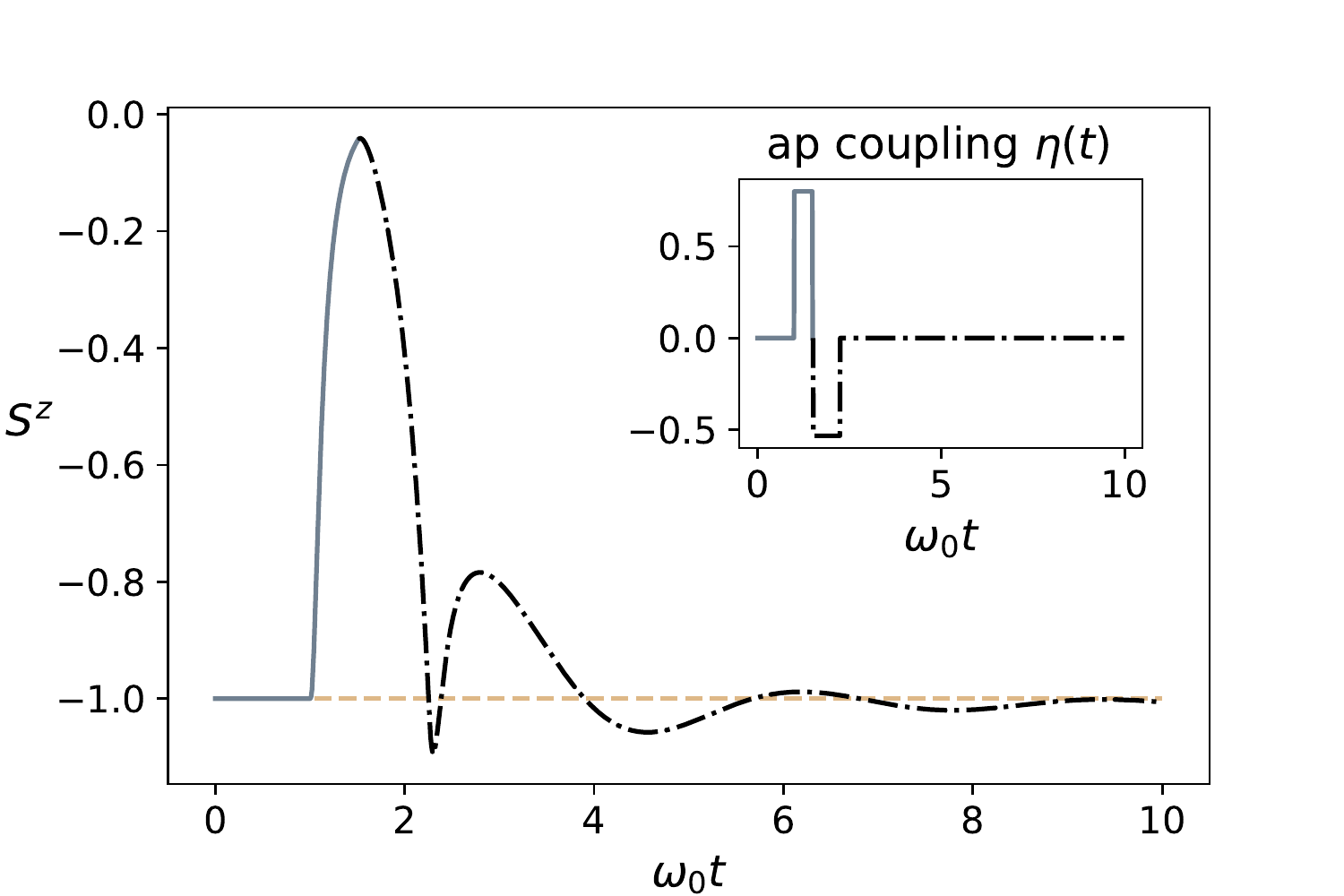}
	\caption{State evolution of the collective neurons under the stimulation (see the inset). The coupling in the storing process is set as 0.8 at the time from 1 to 1.5 ($\omega_0t$) and in the retrieving process set as -0.53 at the time from 1.5 to 2.25 ($\omega_0t$). The system frequency $\omega_0$ is defined as $g$.}
	\label{fig:5}
\end{figure}

To consider an external pulse acting on the neural network, we modify the coupling as a time dependent parameter 
\begin{align}
J(\omega, t) = 2\pi\eta(t)\omega (\frac{\omega}{\omega_c})^{s-1} e^{-\frac{\omega}{\omega_c}},
\end{align} 
meanwhile, at zero temperature, the dissipation and fluctuation coefficients in the master equation are reduced to 
\begin{subequations}
\label{eq:21}
\begin{align}
\kappa(t) &= 2\!\!\int_0^t \!\!ds\!\! \int_0^\infty \!\!\!\!d\omega \eta(s)\omega e^{-\omega/\omega_c}cos[(\omega\!-\!g)(t\!-\!s)], \\
\lambda(t) &= -\!\!\int_0^t \!\!ds \!\!\int_0^\infty \!\!\!\!d\omega \eta(s)\omega e^{-\omega/\omega_c}sin[(\omega\!-\!g)(t\!-\!s)],
\end{align}	
\end{subequations}
and $\widetilde{\kappa}(t) = \widetilde{\lambda}(t) = 0$. Figure \ref{fig:5} shows the dynamics of the collective neural state after stimulation. We apply a simple square pulse with amplitude up to $0.8$ for depolarizing the neural state (see the inset in Fig.~\ref{fig:5}). The gray lines in Fig.~\ref{fig:5} represents time from 0 to 1.5 ($\omega_0t$), and the black-dashed line represents time from 1.5 to 10 ($\omega_0t$), where the duration time is scaled by the system frequency defined as $\omega_0 = 2000$ Hz. Initially, the collective neural state is at polarized state ($S^x$, $S^y$, $S^z$) = (0, 0, -1). At time from 0 to 1 ($\omega_0t$), there is not any environmental noise and the neural system is in the rest state. At time from 1 to 1.5 ($\omega_0t$), we exert a pulse with amplitude 0.8 and the system reaches the depolarized state ($S^x$ ,$S^y$ ,$S^z$) $\approx$ (0, 0, 0). In the retrieving process, we extend the time duration to one and a half compare with the storing process and change the coupling 1.5 times less at time 1.5 to 2.25 ($\omega_0t$), 
then the neural state goes backward, which experiencing the repolarized process. Finally, during the time from 2.25 to 10 ($\omega_0t$), the system gradually returns to the initial rest state.

\begin{figure}[htbp]
	\centering
	\includegraphics[width=1\linewidth]{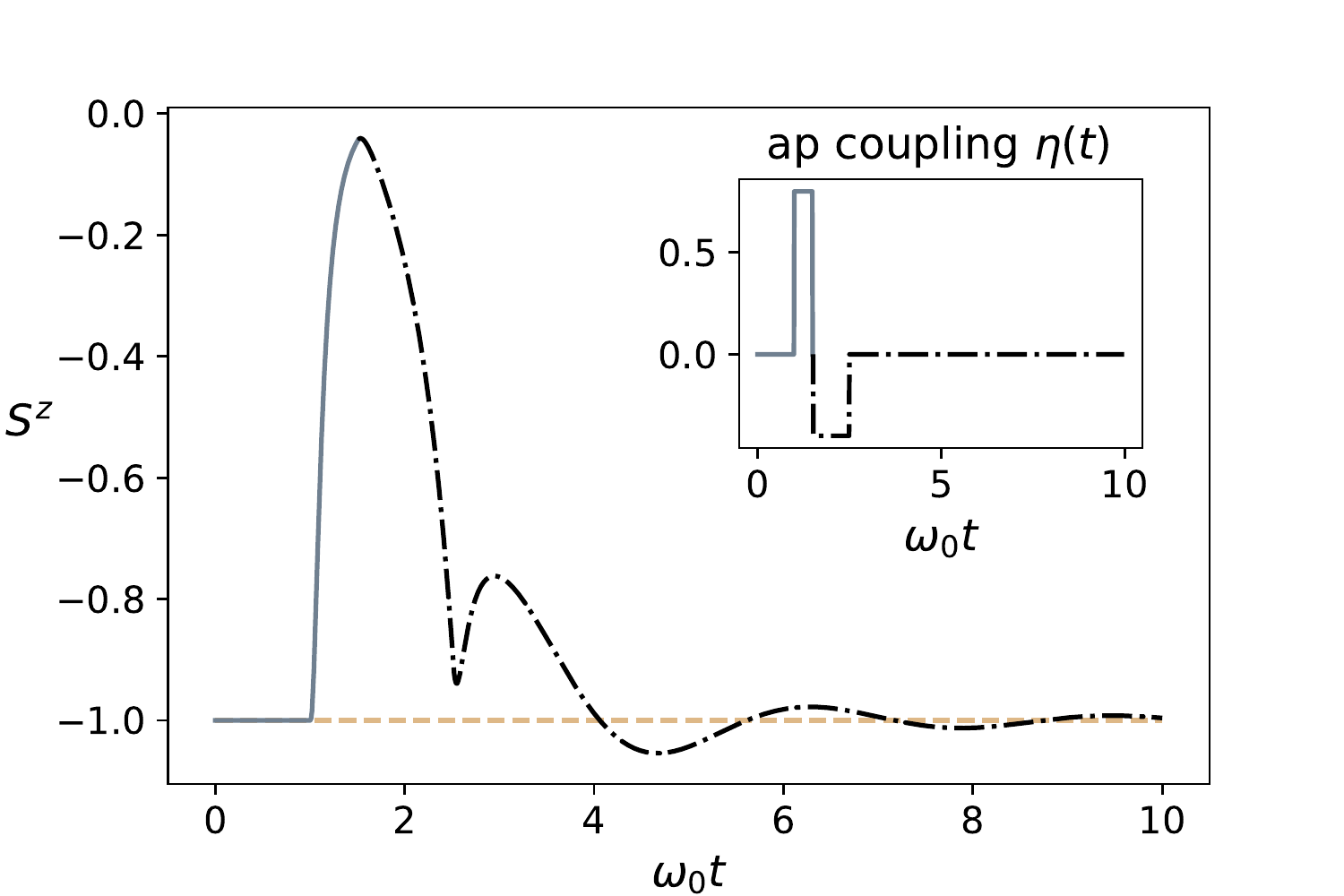}
	\caption{State evolution of the collective neurons with the positive/negative coupling in the same area to the time axis. The negative coupling becomes twice in time and half of the quantity. The coupling in the storing process is set as 0.8 at the time from 1.00 to 1.5 ($\omega_0t$) and in the retrieving process set as -0.4 at the time from 1.5 to 2.5 ($\omega_0t$). The system frequency $\omega_0$ is defined as $g$.}
	\label{fig:6} 
\end{figure}

In Fig.~\ref{fig:6}, we consider the stimulation becoming twice in time and half of the quantity (see the inset of Fig.~\ref{fig:6}). The result shows that the collective neural state can still repolarized to the original rest state. The similar action potential behavior is demonstrated from the collective neural state. This phenomenon shows that the stimulation in the environment activates almost a half of the neuron states so that the collective neural state can reach the depolarized state `$S^Z = 0$'. This is the `depolarization' mechanism of the neural states via the environmental stimulation in our neural model. 

\subsubsection{\label{sec:3_2_2}Collective neural dynamics at room temperature.}

\begin{figure}[htbp]
	\centering	\includegraphics[width=1\linewidth]{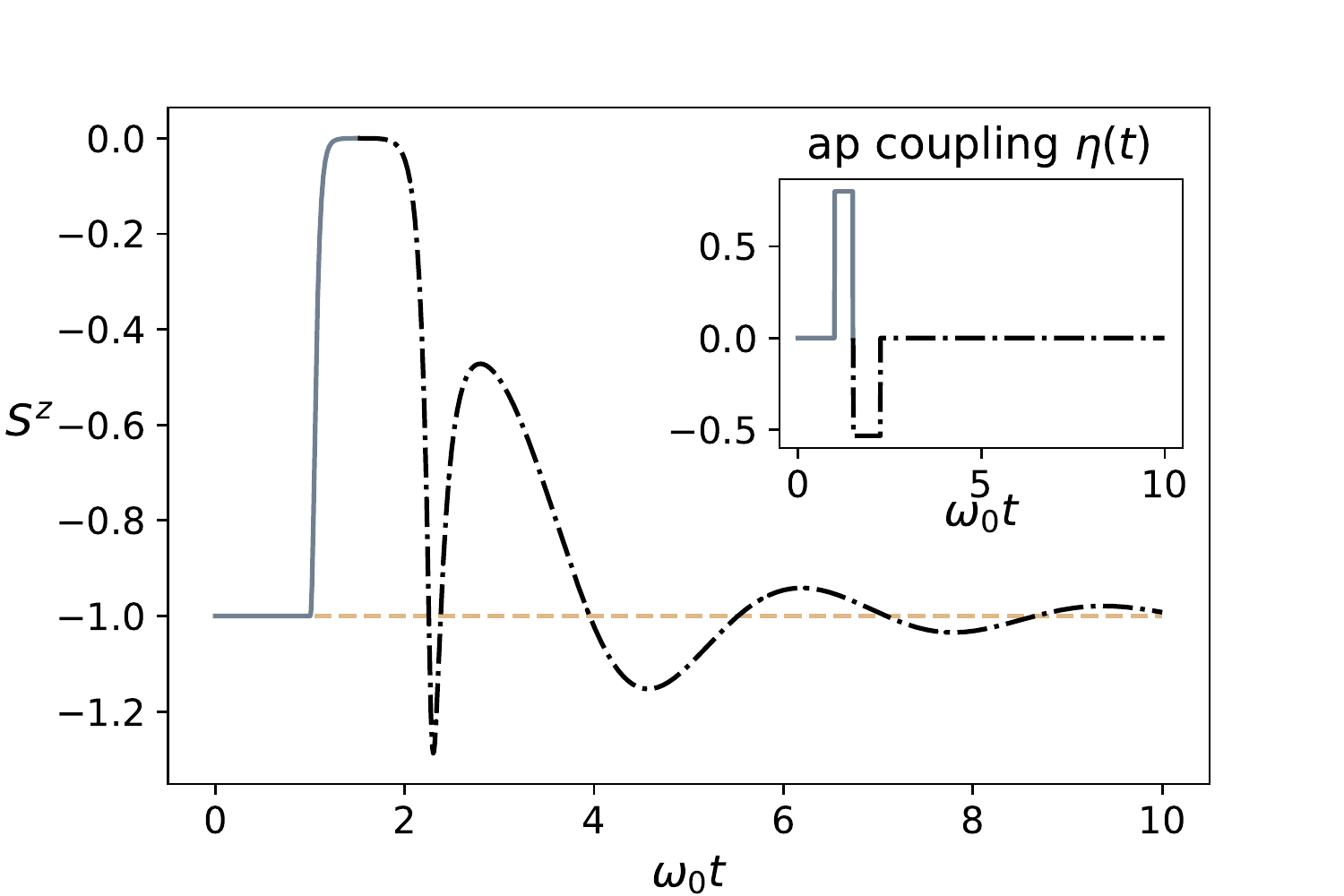}
	\caption{State evolution of the collective neural state for the system considering the environment in the finite temperature ($T = 300 K$). The coupling in the storing process is set as 0.80 at the time from 1.00 to 1.50 ($\omega_0t$) and in the retrieving process set as -0.53 at the time from 1.50 to 2.25 ($\omega_0t$).}
	\label{fig:7}
\end{figure}

\begin{figure}[htbp]
	\centering
	\includegraphics[width=1\linewidth]{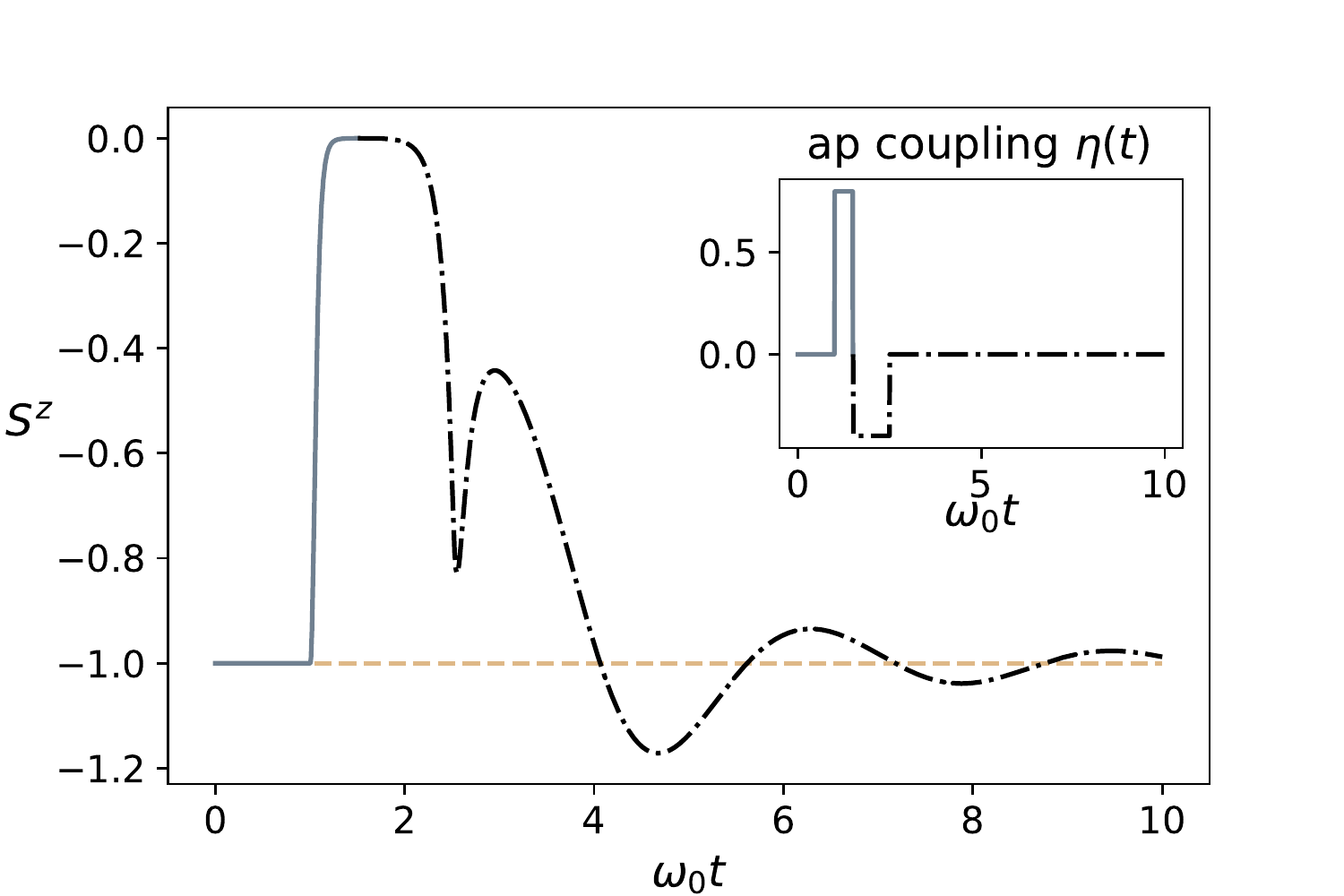}
	\caption{Investigating the positive/negative coupling in the same area to the time axis. Twice of the applying time of the negative coupling as the same in Fig.~\ref{fig:6} to compare with the result in Fig.~\ref{fig:7}. The room temperature is also the same, $T = 300 K$.}
	\label{fig:9}
\end{figure}

However, the zero temperature condition is an ideal case. The bio-system survives within room temperature. The increase of the temperature will give more noise from the environment. If we consider the environment in the room temperature, all the dissipation and fluctuation coefficients of Eq.~(\ref{dfc}) remain. At room temperature ($T\simeq300 K$), the particle distribution in the environment can take the classical Boltzmann distribution ${n}(\omega, T) = e^{-\frac{\hbar\omega}{kT}}$. The state evolution under the room-temperature-environmental effect is shown in Fig.~\ref{fig:7}. In this case, we find that it takes longer time for the state returning back to the initial rest state through the depolarization and repolarization processes due to the environment fluctuation on the collective neural state. This result also shows that no more than a half of the neural states are fired so the maximum amplitude of the collective neural state is a little bit less than $0$ but the temperature effect makes the firing states be more closer to the depolarized state. Furthermore, the condition of the same area also hold with the non-zero-temperature. The result is shown in Fig.~\ref{fig:9} for the same pulse profile in Fig.~\ref{fig:6}.

\section{\label{sec:V}Conclusion}
	In conclusion, we build a neural network model as an open quantum system to understand the randomly coupled neuron-neuron interaction through the coupling with neural environment. We use the master equation approach to study the collective behavior of neurons incorporating the pulse stimulation to demonstrate the action potential. We explore the neuron dynamics at zero temperature and also at room temperature, and find that in both cases, the collective neural states evolve from polarized states (rest states) to the depolarized states and finally back to the initial polarized states under a simplified external pulse driving the environment. Such results show that this simple model can not only catch the expected neuron dynamics but also provide an alternative mechanism and explaining how neurons couple or connect to each other with various wave modes through the complicated neuron surroundings. As the result also shows, the neuron-neuron connections through their surroundings are mainly determined by the spectral density, which characterizes the detailed spectrum structure of the neuron environment. In this work, we only take the simple Ohmic spectral density as an example to simulate the neuron dynamics. The more realistic description of neuron dynamics relies on the spectral density that should be obtained from the spectral measurement of the neural surroundings. Also, the more complete description of the neuron dynamics is given by the neuron firing distribution which is depicted by the reduced density matrix and can be obtained by solving the master equation directly. These remain for the further investigations. 	
 
\begin{acknowledgments}
This work is supported from the Ministry of Science and Technology of Taiwan under
Contract: MOST-108-2112-M-006-009-MY3.
\end{acknowledgments}

\bibliography{references}

\begin{thebibliography}{43}
\expandafter\ifx\csname natexlab\endcsname\relax\def\natexlab#1{#1}\fi
\expandafter\ifx\csname bibnamefont\endcsname\relax
  \def\bibnamefont#1{#1}\fi
\expandafter\ifx\csname bibfnamefont\endcsname\relax
  \def\bibfnamefont#1{#1}\fi
\expandafter\ifx\csname citenamefont\endcsname\relax
  \def\citenamefont#1{#1}\fi
\expandafter\ifx\csname url\endcsname\relax
  \def\url#1{\texttt{#1}}\fi
\expandafter\ifx\csname urlprefix\endcsname\relax\def\urlprefix{URL }\fi
\providecommand{\bibinfo}[2]{#2}
\providecommand{\eprint}[2][]{\url{#2}}

\bibitem[{\citenamefont{Tortora and Derrickson}(2018)}]{tortora2018principles}
\bibinfo{author}{\bibfnamefont{G.~J.} \bibnamefont{Tortora}} \bibnamefont{and}
  \bibinfo{author}{\bibfnamefont{B.~H.} \bibnamefont{Derrickson}},
  \emph{\bibinfo{title}{Principles of anatomy and physiology}}
  (\bibinfo{publisher}{John Wiley \& Sons}, \bibinfo{year}{2018}).

\bibitem[{\citenamefont{LeCun et~al.}(2015)\citenamefont{LeCun, Bengio, and
  Hinton}}]{lecun2015deep}
\bibinfo{author}{\bibfnamefont{Y.}~\bibnamefont{LeCun}},
  \bibinfo{author}{\bibfnamefont{Y.}~\bibnamefont{Bengio}}, \bibnamefont{and}
  \bibinfo{author}{\bibfnamefont{G.}~\bibnamefont{Hinton}},
  \bibinfo{journal}{Nature} \textbf{\bibinfo{volume}{521}},
  \bibinfo{pages}{436} (\bibinfo{year}{2015}).

\bibitem[{\citenamefont{Voulodimos et~al.}(2018)\citenamefont{Voulodimos,
  Doulamis, Doulamis, and Protopapadakis}}]{voulodimos2018deep}
\bibinfo{author}{\bibfnamefont{A.}~\bibnamefont{Voulodimos}},
  \bibinfo{author}{\bibfnamefont{N.}~\bibnamefont{Doulamis}},
  \bibinfo{author}{\bibfnamefont{A.}~\bibnamefont{Doulamis}}, \bibnamefont{and}
  \bibinfo{author}{\bibfnamefont{E.}~\bibnamefont{Protopapadakis}},
  \bibinfo{journal}{Computational Intelligence and Neuroscience}
  \textbf{\bibinfo{volume}{2018}} (\bibinfo{year}{2018}).

\bibitem[{\citenamefont{Nassif et~al.}(2019)\citenamefont{Nassif, Shahin,
  Attili, Azzeh, and Shaalan}}]{nassif2019speech}
\bibinfo{author}{\bibfnamefont{A.~B.} \bibnamefont{Nassif}},
  \bibinfo{author}{\bibfnamefont{I.}~\bibnamefont{Shahin}},
  \bibinfo{author}{\bibfnamefont{I.}~\bibnamefont{Attili}},
  \bibinfo{author}{\bibfnamefont{M.}~\bibnamefont{Azzeh}}, \bibnamefont{and}
  \bibinfo{author}{\bibfnamefont{K.}~\bibnamefont{Shaalan}},
  \bibinfo{journal}{IEEE Access} \textbf{\bibinfo{volume}{7}},
  \bibinfo{pages}{19143} (\bibinfo{year}{2019}).

\bibitem[{\citenamefont{Evgeniou and Pontil}(2004)}]{evgeniou2004regularized}
\bibinfo{author}{\bibfnamefont{T.}~\bibnamefont{Evgeniou}} \bibnamefont{and}
  \bibinfo{author}{\bibfnamefont{M.}~\bibnamefont{Pontil}}, in
  \emph{\bibinfo{booktitle}{Proceedings of the Tenth ACM SIGKDD International
  Conference on Knowledge Discovery and Data Mining}}
  (\bibinfo{publisher}{Association for Computing Machinery},
  \bibinfo{address}{New York, NY, USA}, \bibinfo{year}{2004}), KDD '04, p.
  \bibinfo{pages}{109–117}, ISBN \bibinfo{isbn}{1581138881},
  \urlprefix\url{https://doi.org/10.1145/1014052.1014067}.

\bibitem[{\citenamefont{Xiao et~al.}(2020)\citenamefont{Xiao, Codevilla,
  Gurram, Urfalioglu, and L{\'o}pez}}]{xiao2020multimodal}
\bibinfo{author}{\bibfnamefont{Y.}~\bibnamefont{Xiao}},
  \bibinfo{author}{\bibfnamefont{F.}~\bibnamefont{Codevilla}},
  \bibinfo{author}{\bibfnamefont{A.}~\bibnamefont{Gurram}},
  \bibinfo{author}{\bibfnamefont{O.}~\bibnamefont{Urfalioglu}},
  \bibnamefont{and} \bibinfo{author}{\bibfnamefont{A.~M.}
  \bibnamefont{L{\'o}pez}}, \bibinfo{journal}{IEEE Transactions on Intelligent
  Transportation Systems} pp. \bibinfo{pages}{1--11} (\bibinfo{year}{2020}).

\bibitem[{\citenamefont{Hassabis et~al.}(2017)\citenamefont{Hassabis, Kumaran,
  Summerfield, and Botvinick}}]{hassabis2017neuroscience}
\bibinfo{author}{\bibfnamefont{D.}~\bibnamefont{Hassabis}},
  \bibinfo{author}{\bibfnamefont{D.}~\bibnamefont{Kumaran}},
  \bibinfo{author}{\bibfnamefont{C.}~\bibnamefont{Summerfield}},
  \bibnamefont{and}
  \bibinfo{author}{\bibfnamefont{M.}~\bibnamefont{Botvinick}},
  \bibinfo{journal}{Neuron} \textbf{\bibinfo{volume}{95}}, \bibinfo{pages}{245}
  (\bibinfo{year}{2017}).

\bibitem[{\citenamefont{Turing}(1937)}]{turing1937computable}
\bibinfo{author}{\bibfnamefont{A.~M.} \bibnamefont{Turing}},
  \bibinfo{journal}{Proceedings of the London mathematical society}
  \textbf{\bibinfo{volume}{2}}, \bibinfo{pages}{230} (\bibinfo{year}{1937}).

\bibitem[{\citenamefont{Turing}(2009)}]{turing2009computing}
\bibinfo{author}{\bibfnamefont{A.~M.} \bibnamefont{Turing}}, in
  \emph{\bibinfo{booktitle}{Parsing the Turing Test.}}
  (\bibinfo{publisher}{Springer, Dordrecht.}, \bibinfo{year}{2009}), pp.
  \bibinfo{pages}{23--65}.

\bibitem[{\citenamefont{McCulloch and Pitts}(1943)}]{mcculloch1943logical}
\bibinfo{author}{\bibfnamefont{W.~S.} \bibnamefont{McCulloch}}
  \bibnamefont{and} \bibinfo{author}{\bibfnamefont{W.}~\bibnamefont{Pitts}},
  \bibinfo{journal}{The bulletin of mathematical biophysics}
  \textbf{\bibinfo{volume}{5}}, \bibinfo{pages}{115} (\bibinfo{year}{1943}).

\bibitem[{\citenamefont{Hodgkin and Huxley}(1952)}]{hodgkin1952quantitative}
\bibinfo{author}{\bibfnamefont{A.~L.} \bibnamefont{Hodgkin}} \bibnamefont{and}
  \bibinfo{author}{\bibfnamefont{A.~F.} \bibnamefont{Huxley}},
  \bibinfo{journal}{The Journal of physiology} \textbf{\bibinfo{volume}{117}},
  \bibinfo{pages}{500} (\bibinfo{year}{1952}).

\bibitem[{\citenamefont{Cole and Curtis}(1939)}]{cole1939electric}
\bibinfo{author}{\bibfnamefont{K.~S.} \bibnamefont{Cole}} \bibnamefont{and}
  \bibinfo{author}{\bibfnamefont{H.~J.} \bibnamefont{Curtis}},
  \bibinfo{journal}{The Journal of general physiology}
  \textbf{\bibinfo{volume}{22}}, \bibinfo{pages}{649} (\bibinfo{year}{1939}).

\bibitem[{\citenamefont{Hodgkin and Katz}(1949)}]{hodgkin1949effect}
\bibinfo{author}{\bibfnamefont{A.~L.} \bibnamefont{Hodgkin}} \bibnamefont{and}
  \bibinfo{author}{\bibfnamefont{B.}~\bibnamefont{Katz}}, \bibinfo{journal}{The
  Journal of Physiology} \textbf{\bibinfo{volume}{108}}, \bibinfo{pages}{37}
  (\bibinfo{year}{1949}).

\bibitem[{\citenamefont{Keynes}(1951)}]{keynes1951ionic}
\bibinfo{author}{\bibfnamefont{R.}~\bibnamefont{Keynes}}, \bibinfo{journal}{The
  Journal of physiology} \textbf{\bibinfo{volume}{114}}, \bibinfo{pages}{119}
  (\bibinfo{year}{1951}).

\bibitem[{\citenamefont{Hodgkin}(1971)}]{hodgkin1971conduction}
\bibinfo{author}{\bibfnamefont{A.~L.} \bibnamefont{Hodgkin}},
  \emph{\bibinfo{title}{The Conduction of the Nervous Impulse}}, Sherrington
  lectures (\bibinfo{publisher}{Liverpool University Press},
  \bibinfo{year}{1971}), ISBN \bibinfo{isbn}{9780853230618},
  \urlprefix\url{https://books.google.com.tw/books?id=EuKhAQAACAAJ}.

\bibitem[{\citenamefont{Bean}(2007)}]{bean2007action}
\bibinfo{author}{\bibfnamefont{B.~P.} \bibnamefont{Bean}},
  \bibinfo{journal}{Nature Reviews Neuroscience} \textbf{\bibinfo{volume}{8}},
  \bibinfo{pages}{451} (\bibinfo{year}{2007}).

\bibitem[{\citenamefont{Hebb}(1949)}]{hebb1949organization}
\bibinfo{author}{\bibfnamefont{D.~O.} \bibnamefont{Hebb}}, \bibinfo{journal}{A
  Wiley Book in Clinical Psychology} \textbf{\bibinfo{volume}{62}},
  \bibinfo{pages}{78} (\bibinfo{year}{1949}).

\bibitem[{\citenamefont{Hopfield}(1982)}]{hopfield1982neural}
\bibinfo{author}{\bibfnamefont{J.~J.} \bibnamefont{Hopfield}},
  \bibinfo{journal}{Proceedings of the National Academy of Sciences}
  \textbf{\bibinfo{volume}{79}}, \bibinfo{pages}{2554} (\bibinfo{year}{1982}).

\bibitem[{\citenamefont{Sofroniew and Vinters}(2010)}]{sofroniew2010astrocytes}
\bibinfo{author}{\bibfnamefont{M.~V.} \bibnamefont{Sofroniew}}
  \bibnamefont{and} \bibinfo{author}{\bibfnamefont{H.~V.}
  \bibnamefont{Vinters}}, \bibinfo{journal}{Acta Neuropathologica}
  \textbf{\bibinfo{volume}{119}}, \bibinfo{pages}{7} (\bibinfo{year}{2010}).

\bibitem[{\citenamefont{Kaufmann et~al.}(1989)\citenamefont{Kaufmann, Hanke,
  and Corcia}}]{kaufmann1989ion}
\bibinfo{author}{\bibfnamefont{K.}~\bibnamefont{Kaufmann}},
  \bibinfo{author}{\bibfnamefont{W.}~\bibnamefont{Hanke}}, \bibnamefont{and}
  \bibinfo{author}{\bibfnamefont{A.}~\bibnamefont{Corcia}}
  (\bibinfo{year}{1989}).

\bibitem[{\citenamefont{Pollack}(2001)}]{pollack2001cells}
\bibinfo{author}{\bibfnamefont{G.~H.} \bibnamefont{Pollack}},
  \emph{\bibinfo{title}{Cells, gels and the engines of life: a new, unifying
  approach to cell function}} (\bibinfo{publisher}{Ebner \& Sons Seattle, WA},
  \bibinfo{year}{2001}).

\bibitem[{\citenamefont{Ivanova et~al.}(2003)\citenamefont{Ivanova, Makarov,
  Sch{\"a}ffer, and Heimburg}}]{ivanova2003analyzing}
\bibinfo{author}{\bibfnamefont{V.}~\bibnamefont{Ivanova}},
  \bibinfo{author}{\bibfnamefont{I.}~\bibnamefont{Makarov}},
  \bibinfo{author}{\bibfnamefont{T.}~\bibnamefont{Sch{\"a}ffer}},
  \bibnamefont{and} \bibinfo{author}{\bibfnamefont{T.}~\bibnamefont{Heimburg}},
  \bibinfo{journal}{Biophysical journal} \textbf{\bibinfo{volume}{84}},
  \bibinfo{pages}{2427} (\bibinfo{year}{2003}).

\bibitem[{\citenamefont{Heimburg and Jackson}(2005)}]{heimburg2005soliton}
\bibinfo{author}{\bibfnamefont{T.}~\bibnamefont{Heimburg}} \bibnamefont{and}
  \bibinfo{author}{\bibfnamefont{A.~D.} \bibnamefont{Jackson}},
  \bibinfo{journal}{Proceedings of the National Academy of Sciences}
  \textbf{\bibinfo{volume}{102}}, \bibinfo{pages}{9790} (\bibinfo{year}{2005}).

\bibitem[{\citenamefont{Rose et~al.}(2016)\citenamefont{Rose, Macieszczak,
  Lesanovsky, and Garrahan}}]{rose2016metastability}
\bibinfo{author}{\bibfnamefont{D.~C.} \bibnamefont{Rose}},
  \bibinfo{author}{\bibfnamefont{K.}~\bibnamefont{Macieszczak}},
  \bibinfo{author}{\bibfnamefont{I.}~\bibnamefont{Lesanovsky}},
  \bibnamefont{and} \bibinfo{author}{\bibfnamefont{J.~P.}
  \bibnamefont{Garrahan}}, \bibinfo{journal}{Physical Review E}
  \textbf{\bibinfo{volume}{94}}, \bibinfo{pages}{052132}
  (\bibinfo{year}{2016}).

\bibitem[{\citenamefont{Rotondo et~al.}(2018)\citenamefont{Rotondo, Marcuzzi,
  Garrahan, Lesanovsky, and M{\"u}ller}}]{rotondo2018open}
\bibinfo{author}{\bibfnamefont{P.}~\bibnamefont{Rotondo}},
  \bibinfo{author}{\bibfnamefont{M.}~\bibnamefont{Marcuzzi}},
  \bibinfo{author}{\bibfnamefont{J.~P.} \bibnamefont{Garrahan}},
  \bibinfo{author}{\bibfnamefont{I.}~\bibnamefont{Lesanovsky}},
  \bibnamefont{and}
  \bibinfo{author}{\bibfnamefont{M.}~\bibnamefont{M{\"u}ller}},
  \bibinfo{journal}{Journal of Physics A: Mathematical and Theoretical}
  \textbf{\bibinfo{volume}{51}}, \bibinfo{pages}{115301}
  (\bibinfo{year}{2018}).

\bibitem[{\citenamefont{Damanet et~al.}(2019)\citenamefont{Damanet, Daley, and
  Keeling}}]{damanet2019atom}
\bibinfo{author}{\bibfnamefont{F.}~\bibnamefont{Damanet}},
  \bibinfo{author}{\bibfnamefont{A.~J.} \bibnamefont{Daley}}, \bibnamefont{and}
  \bibinfo{author}{\bibfnamefont{J.}~\bibnamefont{Keeling}},
  \bibinfo{journal}{Physical Review A} \textbf{\bibinfo{volume}{99}},
  \bibinfo{pages}{033845} (\bibinfo{year}{2019}).

\bibitem[{\citenamefont{Glauber}(1963)}]{glauber1963time}
\bibinfo{author}{\bibfnamefont{R.~J.} \bibnamefont{Glauber}},
  \bibinfo{journal}{Journal of mathematical physics}
  \textbf{\bibinfo{volume}{4}}, \bibinfo{pages}{294} (\bibinfo{year}{1963}).

\bibitem[{\citenamefont{Peretto}(1984)}]{peretto1984collective}
\bibinfo{author}{\bibfnamefont{P.}~\bibnamefont{Peretto}},
  \bibinfo{journal}{Biological cybernetics} \textbf{\bibinfo{volume}{50}},
  \bibinfo{pages}{51} (\bibinfo{year}{1984}).

\bibitem[{\citenamefont{Cole and Baker}(1941)}]{cole1941longitudinal}
\bibinfo{author}{\bibfnamefont{K.~S.} \bibnamefont{Cole}} \bibnamefont{and}
  \bibinfo{author}{\bibfnamefont{R.~F.} \bibnamefont{Baker}},
  \bibinfo{journal}{The Journal of general physiology}
  \textbf{\bibinfo{volume}{24}}, \bibinfo{pages}{771} (\bibinfo{year}{1941}).

\bibitem[{\citenamefont{Wang et~al.}(2018)\citenamefont{Wang, Wang, Thow, Lee,
  Peh, Ng, He, Thakor, Chen, and Lee}}]{wang2018inductance}
\bibinfo{author}{\bibfnamefont{H.}~\bibnamefont{Wang}},
  \bibinfo{author}{\bibfnamefont{J.}~\bibnamefont{Wang}},
  \bibinfo{author}{\bibfnamefont{X.~Y.} \bibnamefont{Thow}},
  \bibinfo{author}{\bibfnamefont{S.}~\bibnamefont{Lee}},
  \bibinfo{author}{\bibfnamefont{W.~Y.~X.} \bibnamefont{Peh}},
  \bibinfo{author}{\bibfnamefont{K.~A.} \bibnamefont{Ng}},
  \bibinfo{author}{\bibfnamefont{T.}~\bibnamefont{He}},
  \bibinfo{author}{\bibfnamefont{N.~V.} \bibnamefont{Thakor}},
  \bibinfo{author}{\bibfnamefont{C.-H.} \bibnamefont{Chen}}, \bibnamefont{and}
  \bibinfo{author}{\bibfnamefont{C.}~\bibnamefont{Lee}},
  \bibinfo{journal}{bioRxiv} p. \bibinfo{pages}{343905} (\bibinfo{year}{2018}).

\bibitem[{\citenamefont{Feynman and Vernon~Jr}(2000)}]{feynman2000theory}
\bibinfo{author}{\bibfnamefont{R.~P.} \bibnamefont{Feynman}} \bibnamefont{and}
  \bibinfo{author}{\bibfnamefont{F.~L.} \bibnamefont{Vernon~Jr}},
  \bibinfo{journal}{Annals of physics} \textbf{\bibinfo{volume}{281}},
  \bibinfo{pages}{547} (\bibinfo{year}{2000}).

\bibitem[{\citenamefont{Rosenblatt}(1961)}]{rosenblatt1961principles}
\bibinfo{author}{\bibfnamefont{F.}~\bibnamefont{Rosenblatt}},
  \bibinfo{type}{Tech. Rep.}, \bibinfo{institution}{Cornell Aeronautical Lab
  Inc Buffalo NY} (\bibinfo{year}{1961}).

\bibitem[{\citenamefont{Cooper et~al.}(1979)\citenamefont{Cooper, Liberman, and
  Oja}}]{cooper1979theory}
\bibinfo{author}{\bibfnamefont{L.~N.} \bibnamefont{Cooper}},
  \bibinfo{author}{\bibfnamefont{F.}~\bibnamefont{Liberman}}, \bibnamefont{and}
  \bibinfo{author}{\bibfnamefont{E.}~\bibnamefont{Oja}},
  \bibinfo{journal}{Biological cybernetics} \textbf{\bibinfo{volume}{33}},
  \bibinfo{pages}{9} (\bibinfo{year}{1979}).

\bibitem[{\citenamefont{Longuet-Higgins}(1968{\natexlab{a}})}]{longuet1968non}
\bibinfo{author}{\bibfnamefont{H.~C.} \bibnamefont{Longuet-Higgins}},
  \bibinfo{journal}{Proceedings of the Royal Society of London. Series B.
  Biological Sciences} \textbf{\bibinfo{volume}{171}}, \bibinfo{pages}{327}
  (\bibinfo{year}{1968}{\natexlab{a}}).

\bibitem[{\citenamefont{Longuet-Higgins}(1968{\natexlab{b}})}]{longuet1968holographic}
\bibinfo{author}{\bibfnamefont{H.~C.} \bibnamefont{Longuet-Higgins}},
  \bibinfo{journal}{Nature} \textbf{\bibinfo{volume}{217}},
  \bibinfo{pages}{104} (\bibinfo{year}{1968}{\natexlab{b}}).

\bibitem[{\citenamefont{Kohonen}(2012)}]{kohonen2012associative}
\bibinfo{author}{\bibfnamefont{T.}~\bibnamefont{Kohonen}},
  \emph{\bibinfo{title}{Associative memory: A system-theoretical approach}},
  vol.~\bibinfo{volume}{17} (\bibinfo{publisher}{Springer Science \& Business
  Media}, \bibinfo{year}{2012}).

\bibitem[{\citenamefont{Amari}(1977)}]{amari1977neural}
\bibinfo{author}{\bibfnamefont{S.-I.} \bibnamefont{Amari}},
  \bibinfo{journal}{Biological cybernetics} \textbf{\bibinfo{volume}{26}},
  \bibinfo{pages}{175} (\bibinfo{year}{1977}).

\bibitem[{\citenamefont{Amari and Takeuchi}(1978)}]{amari1978mathematical}
\bibinfo{author}{\bibfnamefont{S.-i.} \bibnamefont{Amari}} \bibnamefont{and}
  \bibinfo{author}{\bibfnamefont{A.}~\bibnamefont{Takeuchi}},
  \bibinfo{journal}{Biological Cybernetics} \textbf{\bibinfo{volume}{29}},
  \bibinfo{pages}{127} (\bibinfo{year}{1978}).

\bibitem[{\citenamefont{Little}(1974)}]{little1974existence}
\bibinfo{author}{\bibfnamefont{W.~A.} \bibnamefont{Little}},
  \bibinfo{journal}{Mathematical biosciences} \textbf{\bibinfo{volume}{19}},
  \bibinfo{pages}{101} (\bibinfo{year}{1974}).

\bibitem[{\citenamefont{Amit et~al.}(1985)\citenamefont{Amit, Gutfreund, and
  Sompolinsky}}]{amit1985spin}
\bibinfo{author}{\bibfnamefont{D.~J.} \bibnamefont{Amit}},
  \bibinfo{author}{\bibfnamefont{H.}~\bibnamefont{Gutfreund}},
  \bibnamefont{and}
  \bibinfo{author}{\bibfnamefont{H.}~\bibnamefont{Sompolinsky}},
  \bibinfo{journal}{Physical Review A} \textbf{\bibinfo{volume}{32}},
  \bibinfo{pages}{1007} (\bibinfo{year}{1985}).

\bibitem[{\citenamefont{Von~Neumann}(2018)}]{von2018mathematical}
\bibinfo{author}{\bibfnamefont{J.}~\bibnamefont{Von~Neumann}},
  \emph{\bibinfo{title}{Mathematical foundations of quantum mechanics: New
  edition}} (\bibinfo{publisher}{Princeton university press},
  \bibinfo{year}{2018}).

\bibitem[{\citenamefont{Breuer and Petruccione}(2002)}]{breuer2002theory}
\bibinfo{author}{\bibfnamefont{H.-P.} \bibnamefont{Breuer}} \bibnamefont{and}
  \bibinfo{author}{\bibfnamefont{F.}~\bibnamefont{Petruccione}},
  \emph{\bibinfo{title}{The theory of open quantum systems}}
  (\bibinfo{publisher}{Oxford University Press on Demand},
  \bibinfo{year}{2002}).

\bibitem[{\citenamefont{Leggett et~al.}(1987)\citenamefont{Leggett,
  Chakravarty, Dorsey, Fisher, Garg, and Zwerger}}]{leggett1987dynamics}
\bibinfo{author}{\bibfnamefont{A.~J.} \bibnamefont{Leggett}},
  \bibinfo{author}{\bibfnamefont{S.}~\bibnamefont{Chakravarty}},
  \bibinfo{author}{\bibfnamefont{A.~T.} \bibnamefont{Dorsey}},
  \bibinfo{author}{\bibfnamefont{M.~P.} \bibnamefont{Fisher}},
  \bibinfo{author}{\bibfnamefont{A.}~\bibnamefont{Garg}}, \bibnamefont{and}
  \bibinfo{author}{\bibfnamefont{W.}~\bibnamefont{Zwerger}},
  \bibinfo{journal}{Reviews of Modern Physics} \textbf{\bibinfo{volume}{59}},
  \bibinfo{pages}{1} (\bibinfo{year}{1987}).

\end{thebibliography}

\end{document}